\documentclass[twocolumn,english,notitlepage,superscriptaddress]{revtex4-1}
\usepackage[T1]{fontenc}
\usepackage[latin9]{inputenc}
\usepackage{bm}
\usepackage{amsmath}
\usepackage{amssymb}
\usepackage{graphicx}
\usepackage{babel}
\begin{document}

\title{Detection of anisotropic particles in levitated optomechanics}

\author{Marko Toro\v{s}}
\email{m.toros@soton.ac.uk}

\affiliation{Department of Physics and Astronomy, University of Southampton, SO17
1BJ, United Kingdom}

\author{Muddassar Rashid}

\affiliation{Department of Physics and Astronomy, University of Southampton, SO17
1BJ, United Kingdom}

\author{Hendrik Ulbricht}
\email{h.ulbricht@soton.ac.uk }

\affiliation{Department of Physics and Astronomy, University of Southampton, SO17
1BJ, United Kingdom}
\begin{abstract}
We discuss the detection of an anisotropic particle trapped by
an elliptically polarized focused Gaussian laser beam. We
obtain the full rotational and translational dynamics, as well as, the measured photo-current in a general-dyne detection. As an example, we discuss a toy model of homodyne detection, which captures the main features typically found in experimental setups.
\end{abstract}
\maketitle

\section{Introduction}

Nanoparticles in optical traps are becoming increasingly interesting
as they hold the promise of exploring quantum features at novel
scales.  Typical nanoparticles of mass $10^{-21}$ - $10^{-18}$ kg will
push the classical-quantum boundary of exploration into the mesoscopic regime, improving by several orders on the mass $10^{-23}$ kg, which is the
most massive object to have been shown to exhibit quantum
interference~\cite{eibenberger2013matter}.  Consequently, such systems
can be used to test the superposition principle \cite{bateman2014}, as
well as, for the detection of small forces ~\cite{chang2010cavity,
  romero2010toward, ranjit2016zeptonewton, bassi2013models,
  hempston2017force}.

The most direct approach to reach the quantum regime is to cool the system
to the ground state in high vacuum~\cite{li2010measurement,
  jain2016direct, millen2015cavity,
  kiesel2013cavity,asenbaum2013cavity, vovrosh2017parametric,
  setter2017real}.  This endeavour, which has proven to be
non-trivial, has lead to a detailed analysis of the forces involved,
namely light-matter interaction and gas
collisions~\cite{jain2016direct}, as
  well as gravity~\cite{hebestreit2017sensing}.  The nanoparticle is
often a small homogeneous sphere, which can be modelled as a polarizable
point particle in a harmonic trap, leading to a distinct harmonic motion for each of
the three translational degrees of freedom.

However, it has been recently shown that a non-spherical nanoparticle, of a prefabricated shape, leads to interesting rotational
~\cite{kuhn2015cavity,hoang2016torsional,kane2010levitated,arita2013laser,delord2017electron,coppock2016phase}
and librational motion \cite{hoang2016torsional}. Furthermore, these
investigations have sparked the discussion of some novel ideas in
levitated optomechanics, namely force-sensing using spinning
objects~\cite{kuhn2017s,kuhn2017full,rashid2018precession,Manjavacas2017}, reaching the ground state of librational
motion~\cite{zhong2017shot}, and the generation of quantum superpositions of such rotational degrees of freedom~\cite{stickler2018orientational}.  Such anisotropic objects have three
translational, as well as, three rotational degrees of freedom, where
the latter ones, are commonly known as the rigid rotor. These have
been studied extensively in both
classical~\cite{goldstein2011classical,arnol2013mathematical} and
quantum mechanics~\cite{casimir1931rotation,biedenharn1984angular}. However, only recently
has the investigation of the rotational degrees been extended to open quantum
systems~\cite{papendell2017quantum,liu2017coupling,stickler2016rotranslational,
stickler2015molecular,stickler2016rotranslational,stickler2017rotational,stickler2016spatio}.

To realise such novel experiments, it is imperative to gain a detailed
understanding of the rich dynamics a nanoparticle can
exhibit: these motions can only be extracted through measurement~\cite{jones2009rotation}. It is thus necessary to consider, not only the system dynamics, but also the detection method, i.e. the measurement apparatus, to give a complete description of an experiment. This can be already important for classical systems, where a measurement using a physical procedure will generally perturb a small system, but the two become even more intertwined in the quantum case, where each measurement will change the system and thus also its subsequent evolution. Moreover, when the system has several degrees of freedom, extracting the motion of a particular degree of freedom becomes a non-trivial exercise: the majority of the detection schemes rely on scattering from the trapped particle which invariably carries information on translational, rotational and librational motions, first coupled in a complicated motion, and then mapped  into a scalar signal at the detector. 

In this paper, building on the previous work, we investigate the
rotational and translational (ro-translational) motion of such systems, namely that of an
anisotropic polarizable particle in an optical trap. We will
consider light-matter interactions, namely the quantum analogue of the
gradient, scattering forces and torques. Specifically, we will discuss the case of an
elliptically polarized Gaussian beam, from which one can also recover
the linear and circular polarizations as limiting cases. In addition, we consider, particle-gas
collisions, modelled by extending the Caldeira-Leggett model to
ro-translations. 

The purpose of this work is twofold. The first goal is to give a detailed
description of the rotational and translational motion under continuous
monitoring. The second is to obtain the formula for the photo-current
in a general dyne detection. This will open the door for the application
of state estimation and manipulation techniques in ro-translational
optomechanics already developed for other quantum systems~\cite{wiseman2009quantum,jacobs2014quantum}. 

This paper is structured as follows. In Sec.~\ref{sec:description-of-the}
we describe the optomechanical system subject to light-matter interactions
and gas collisions. In Sec.~\ref{sec:Dynamics} we then obtain the
quantum dynamics with and without laser monitoring. In Sec.~\ref{sec:dyne}
we discuss the general dyne detection. In addition, we consider a
toy model of homodyne detection, which captures the main features
of typical experimental setups with mirrors and lenses. We write the
conclusions in Sec.~\ref{sec:Conclusion}.

\section{description of the system\label{sec:description-of-the}}

\subsection{Experimental setup }

\begin{figure*}
  \includegraphics[width=0.75\textwidth]{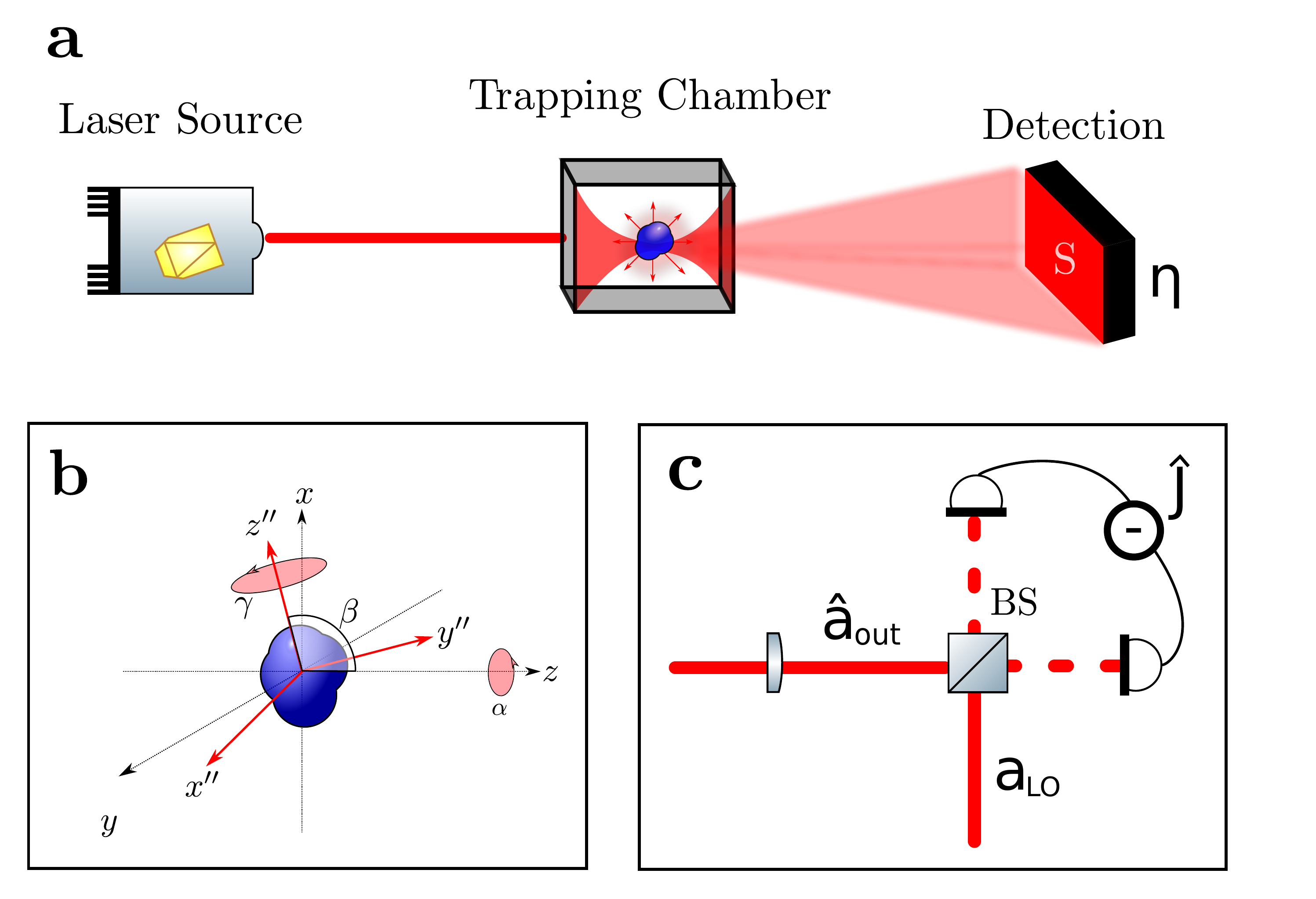}\caption{(a) An
    incoming coherent light beam creates an optical trap. A fraction
    of the photons is scattered, and a fraction of these is then
    recorded by the detector: the surface  and efficiency of the detector are denoted
    by $S$ and $\eta$, respectively. (b) The laboratory axis are denoted by $x$, $y$, $z$ letters,
    while the body-frame axis are denoted by the $x''$, $y''$, $z''$
    letters.  The relation between the two frames is parametrized by
    the Euler angles $\alpha$, $\beta$ and $\gamma$ in z-y'-z''  convention. $\alpha$ denotes the angle of rotation about the
    laboratory $z$ axis (from $x$ towards $y$). $\beta$ is the angle
    between the laboratory $z$ axis and the body $z''$ axis (rotated
    about the $y'$ axis, i.e. the $y$ axis after it has been rotated
    by $\alpha$ about the $z$ axis; from $z$ towards $x$). $\gamma$
    denotes the angle of rotation about the body frame $z''$ axis
    (from $x''$ towards $y''$).  
    (c) Homodyne detection setup. $\hat{a}_\text{out}$ and $a_\text{LO}$ denote the signal (output operator), and the local oscillator (a complex number), respectively. These enter as inputs to the beam splitter, which we denote by BS. The outputs then get subtracted to obtain the measured photo-current $\hat{J}$ in the Heisenberg picture. We denote the corresponding photo-current in the Schr\"{o}dinger picture by $J$ (see Sec.~\ref{sec:dyne}).
    \label{fig:setup}}
\end{figure*}

We consider the experimental setup of an optically levitated particle (see Fig.~\ref{fig:setup}(a)). In a nutshell, a laser light is used to create
an intense focal region inside a trapping chamber (vacuum chamber): once the particle is trapped at
the focus, it will Rayleigh scatter light, which is collected and directed towards a detector. In this paper, we restrict the analysis to experimental situations that can be adequately modelled by a considering a quantization of the electromagnetic field in free space. In general, to model a cavity experiment, one would need to impose appropriate boundary conditions on the electromagnetic field, and repeat the analysis. However, some cavity experiments, e.g. a lossy cavity,  can still be, at least in first approximation, described by the present analysis. In this section
we briefly introduce the main features of this type of experiments using notions from classical electromagnetism and mechanics. We discuss in detail their quantum counter-parts in the following sections.

We first discuss light-matter interactions. The incoming tightly focused light beam with a Gaussian profile creates
an optical trap, which traps a nanoparticle near its focus point.
This corresponds classically to the gradient force and torque. Moreover, the
incoming light beam carries also linear and angular momentum. The
linear momentum creates a radiation pressure scattering force which
displaces the particle along the $z$ axis, while the
angular momentum carried by the photons is transferred to the particle,
which starts to rotate, i.e. spinning. 

We next discuss collisions with the surrounding gas, which is a source of friction. Specifically, the gas of particles acts as a bath for the translational and rotational motions. In the simplest case we expect the particle to eventually reach an out of equilibrium steady state with the surrounding gas: the laser continuously transfers energy to the particle, which is then dissipated into the gas. This results in a specific variance of the translational and librational
degrees of freedom, or, in the case of spinning, an asymptotic angular frequency. 

Both photon scattering and gas collisions are a source of diffusion: each random collision, either with a photon or with a gas particle, makes the particle recoil. Loosely speaking, the net effect of these collisions is a stochastic trajectory of the particle state (monitoring by the environment with unit efficiency). In addition, the interaction with photons, as well as with gas particles, couples the rotational and translational motion: only in some limiting cases the motions decouple.

There is however an important difference between photon scattering and gas collisions. Suppose that the characteristic length of the optically levitated particle is $r_s$, denote the photon wavelength by $\lambda$, and the wavelength associated to a gas particle by $\lambda_g\approx \frac{2 \pi\hbar}{\sqrt{2 m_g k_b T}}$, where  $T$ is the gas temperature, $m_g$ is the mass of a gas particle, and $k_b$ is Boltzman's constant. For photon scattering we are in the long wavelength limit, while for gas collisions, for temperatures above $\sim 1\text{mK}$, we are in the short wavelength limit, i.e. $\lambda_g<r_s<\lambda$. Thus we will model the optically levitated particle in two different ways: on the one hand, for photon scattering, we can approximate it as an anisotropic particle with six degrees of freedom, while, on the other hand, for gas collisions, we will model it initially as a many-body system. However, under some simplifying assumption, e.g. rigid body, the latter will also reduce to the anisotropic particle model with six degrees of freedom.

\subsection{Free Hamiltonian}\label{sec:fh}
We model the optically levitated system as an anisotropic polarizable particle with six degrees of freedom, i.e. three translational and three rotational. We denote the position and momentum operators by
$\hat{\bm{r}}=(\hat{x},\hat{y},\hat{z})^{\top}$ and $\hat{\bm{p}}=(\hat{p}_{x},\hat{p}_{y},\hat{p}_{z})^{\top}$,
respectively, the angle operator by $\hat{\mathbf{\mathbf{\mathbf{\boldsymbol{\phi}}}}}=(\hat{\alpha},\hat{\beta},\hat{\gamma}){}^{\top}$,
where the three operators denote the quantized Euler angles in the
$z$-$y'$-$z''$ convention, and the corresponding (angle) momentum operator
by $\hat{\bm{\pi}}=(\hat{\pi}_{\alpha},\hat{\pi}_{\beta},\hat{\pi}_{\gamma})^{\top}$.

We consider the free Hamiltonian for translational and rotational
degrees of freedom:

\begin{equation}
\hat{H}_{\text{free}}=\frac{\hat{\bm{p}}^{\top}\hat{\bm{p}}}{2M}+\frac{\hat{\mathbf{\bm{\pi}}}^{\top}\hat{N}^{-1}\hat{F}I^{-1}\hat{F}^{\top}(\hat{N}^{\top}){}^{-1}\hat{\mathbf{\bm{\pi}}}}{2},\label{eq:free}
\end{equation}
where $M$ is the mass of the system, $I=\text{diag}(I_{1},I_{2},I_{3})$
is the moment of inertia tensor in the principal axis (the body frame),
$\hat{F}(\hat{\mathbf{\mathbf{\mathbf{\boldsymbol{\phi}}}}})=F_{\text{z}}(\hat{\alpha})F_{\text{y'}}(\hat{\beta})F_{\text{z''}}(\hat{\gamma})$
is the Euler parametrization of a generic rotation, $F_{x}$ denotes
a rotation about the $x$-axis (here $x$ denotes a generic axis),
and $\hat{N}(\hat{\mathbf{\mathbf{\mathbf{\boldsymbol{\phi}}}}})$
is the matrix that maps $\dot{\hat{\mathbf{\mathbf{\mathbf{\boldsymbol{\phi}}}}}}$
to the angular frequency $\hat{\boldsymbol{\omega}}$ in the laboratory
frame, i.e. $\hat{\boldsymbol{\omega}}=\hat{N}(\hat{\boldsymbol{\mathbf{\mathbf{\mathbf{\phi}}}}})\dot{\hat{\mathbf{\boldsymbol{\phi}}}}$ 
(see Fig.~\ref{fig:setup}(b)).

\subsection{Light-matter coupling}
The total electric field $\hat{\bm{E}}$ induces a dipole proportional to $\propto\hat{\chi} \hat{\bm{E}}$, where $\hat{\chi}$ is the susceptibility tensor of the trapped particle, and we suppose that this induced field is coupled with $\bm{\hat{E}}$ by the usual dielectric coupling, i.e. $\propto \hat{\bm{E}}^\top \hat{\chi} \hat{\bm{E}}$. Specifically, we start from the following interaction Hamiltonian:
\begin{equation}
\hat{H}_{\text{int}}=-\frac{1}{2}V\epsilon_{0}\hat{\bm{E}}^{\top}\hat{\chi}\hat{\bm{E}},\label{eq:coupling}
\end{equation}
where $\hat{\bm{E}}(\hat{\bm{r}})$ is the total electric field, $\epsilon_{0}$
is the electric permittivity of free space,  $\hat{\chi}=\hat{F}\chi\hat{F}^{\top}$, $\chi=\text{diag}(\chi_{1},\chi_{2},\chi_{3})$
is the electric susceptibility tensor in the body frame, and $V$ is the volume of the nanoparticle. We
assume that $\chi_{j}$ are $\mathbb{R}$-valued, i.e. we consider only photon scattering, neglecting absorption and emission. 

The total electric field is given by
\begin{equation}
\hat{\bm{E}}=\bm{\hat{E}}_{d}+\bm{\hat{E}}_{f}\label{eq:total_field}
\end{equation}
where $\bm{\hat{E}}_{d}$ is the field that generates the optical
trap and $\hat{\bm{E}}_{f}$ denotes the free electromagnetic field. Loosely speaking, one can think of a single incoming photon travelling in empty space (associated to the $\hat{\bf{E}}_d$ field), that at the nanoparticle location changes to an outgoing photon (associated to either the $\hat{\bf{E}}_d$ or the $\hat{\bf{E}}_f$ field).  This way of separating the electrical field in two terms is reminiscent of the double counting of the output modes in cavity QED~\cite{dutra2005cavity,romero2011optically}. 

Specifically, we consider
\begin{equation}
\bm{\hat{E}}_{d}=iE_{0}(\mathbf{\boldsymbol{\mathbf{\bm{\epsilon}}}}_{d}\hat{u}\hat{a}-\boldsymbol{\mathbf{\bm{\epsilon}}}_{d}^{*}\hat{u}^{*}\hat{a}^{\dagger}),\label{eq:ed}
\end{equation}
where $E_{0}$ is the amplitude of the field, $\boldsymbol{\boldsymbol{\mathbf{\epsilon}}}_{d}$
is the polarization vector, $\hat{u}(\hat{\mathbf{r}})$ is the mode
function, and $\hat{a}$ ($\hat{a}^{\dagger}$) the corresponding annihilation (creation)
operator. Moreover, we will consider the case of elliptical polarization
\begin{equation}
\mathbf{\boldsymbol{\epsilon}}_{d}=\frac{1}{\sqrt{b_{x}^{2}+b_{y}^{2}}}(b_{x},ib_{y},0)^{\top},\label{eq:elliptical}
\end{equation}
where $b_{x}$, $b_{y}$ are $\mathbb{R}$-valued, and $\boldsymbol{\epsilon}_{d}^{*\top}\boldsymbol{\epsilon}_{d}=1$.
More generally, in particular going beyond the paraxial approximation,
one could consider also the case $\boldsymbol{\epsilon}_{d}=\boldsymbol{\epsilon}_{d}(\hat{\mathbf{r}})$.

The free electromagnetic field, which forms a bath, is given by:
\begin{equation}
\bm{\hat{E}}_{f}=i\sum_{\bm{k},\nu}\sqrt{\frac{\hbar\omega_{k}}{2V_{q}\epsilon_{0}}}\left(\bm{\epsilon}_{\bm{k},\nu}\hat{a}_{\bm{k},\nu}e^{i\bm{k}\cdot\hat{\bm{r}}}-\bm{\epsilon}_{\bm{k},\nu}^{*}\hat{a}_{\bm{k},\nu}^{\dagger}e^{-i\bm{k}\cdot\hat{\bm{r}}}\right)\label{eq:ef}
\end{equation}
where $\hat{a}_{\bm{k},\nu}$ ($\hat{a}_{\bm{k},\nu}^{\dagger}$) is
the annihilation (creation) operator, $\bm{\epsilon}_{\bm{k},\nu}$ is the polarization vector, $\bm{k}$
is the wave-vector, $\nu$ denotes the two independent polarizations,
$\omega_{k}=ck$, and $k=\vert\bm{k}\vert$.  
The quantization volume $V_{q}$ is determined the boundaries of the experimental setup~\cite{walls2007quantum}, e.g. $V_{q}=L^3$ with $L$ the size of a box. In case of a cavity system, the boundaries of the problem can be taken as the physical boundaries of the cavity, while for a system in free space the boundaries are at spatial infinity. This latter situation could also be applicable, in first approximation, to a system confined to a large or lossy cavity, i.e. whenever the field description given by Eq.~\eqref{eq:ef} in the limit $V_{q}\rightarrow \infty$ is sufficient. In this paper, we restrict to this latter case of free space quantization, i.e. we consider the continuum limit by making the formal replacements $\sum_{\bm{k},\nu}\rightarrow\frac{V_{q}}{(2\pi)^{3}}\int d\bm{k}$ and  ${\sqrt{V_q}}\hat{a}_{\bf{k},\nu} \rightarrow \hat{a}_{\bf{k},\nu}$. 
Note also that $\bm{\epsilon}_{\bm{k},\nu}=\bm{\epsilon}_{\bm{n},\nu}$
and $\bm{\epsilon}_{\bm{n},\nu}^{*\top}\bm{\epsilon}_{\bm{n},\nu}=1$,
where $\bm{n}$ is a unit vector in the direction of $\bm{k}$, i.e.
$\bm{k}=k\bm{n}$. For more details about the decomposition in Eq.~(\ref{eq:ef})
see Appendix \ref{sec:Polarization-of-scattered}. In the following
we will also use the completeness relation:
\begin{equation}
\sum_{\nu}(\bm{\epsilon}_{\bm{\bm{n}},\nu})_{i}(\bm{\epsilon}_{\bm{\bm{n}},\nu}^{*})_{j}=\delta_{ij}-\bm{n}_{i}\bm{n}_{j}.\label{eq:cr}
\end{equation}
We consider the usual Hamiltonian contribution of the
free electromagnetic field:
\begin{equation}
H_{f}=\sum_{\nu}\int \frac{d\bm{k}}{(2\pi)^3}\hbar\omega_{k}\hat{a}_{\bm{k},\nu}^{\dagger}\hat{a}_{\bm{k},\nu}\label{eq:Hef}
\end{equation}

We now use Eq.~(\ref{eq:total_field}) in Eq.~(\ref{eq:coupling})
from which we obtain two main contributions: the term $\propto\bm{\hat{E}}_{d}^{\top}\hat{\chi}\bm{\hat{E}}_{d}$,
which gives rise to the unitary dynamics, and the term $\propto\bm{\hat{E}}_{f}^{\top}\hat{\chi}\bm{\hat{E}}_{d}$,
which gives rise to the non-unitary dynamics, while we neglect $\propto\bm{\hat{E}}_{f}^{\top}\hat{\chi}\bm{\hat{E}}_{f}$,
as we assume that the free-field modes are initially empty. Classically
these correspond to the gradient and radiation pressure terms, respectively: we now discuss each of these separately. 

\subsubsection{Gradient terms}\label{ssgt}

We consider the term $\propto\bm{\hat{E}}_{d}^{\top}\hat{\chi}\bm{\hat{E}}_{d}$,
where $\bm{\hat{E}}_{d}$ is given in Eq.~(\ref{eq:ed}). Specifically,
from Eqs.~(\ref{eq:coupling})-(\ref{eq:elliptical}), making the
rotating wave approximation (we take the time-average of optical fields, which we assume to oscillate much faster than the typical nanoparticle frequency), we obtain the gradient potential:
\begin{equation}
\hat{H}_{\text{grad}}=-\boldsymbol{\epsilon}_{0}VE_{0}^{2}\vert\hat{u}\vert^{2}(\mathbf{\boldsymbol{\mathbf{\epsilon}}}_{d}^{*})^{\top}\hat{\chi}\mathbf{\boldsymbol{\epsilon}}_{d}\hat{a}^{\dagger}\hat{a},\label{eq:gradient}
\end{equation}
where 
\begin{alignat}{1}
(\boldsymbol{\mathbf{\epsilon}}_{d}^{*})^{\top}\hat{\chi}\mathbf{\boldsymbol{\epsilon}}_{d}&=  b_{x}^{2}\bigg[\text{\ensuremath{\chi}}_{1}(\cos(\hat{\alpha})\cos(\hat{\beta})\cos(\hat{\gamma})-\sin(\hat{\alpha})\sin(\hat{\gamma}))^{2}\nonumber \\
 &\;\;\;\;+\text{\ensuremath{\chi}}_{2}(\cos(\hat{\alpha})\cos(\hat{\beta})\sin(\hat{\gamma})+\sin(\hat{\alpha})\cos(\gamma))^{2}\nonumber\\
 &\;\;\;\;+\text{\ensuremath{\chi}}_{3}\cos^{2}(\alpha)\sin^{2}(\beta)\bigg]\nonumber \\
 & +b_{y}^{2}\bigg[\text{\ensuremath{\chi}}_{1}(\sin(\hat{\alpha})\cos(\hat{\beta})\cos(\hat{\gamma})+\cos(\hat{\alpha})\sin(\hat{\gamma}))^{2}\nonumber \\
 &\;\;\;\;+\text{\ensuremath{\chi}}_{2}(\cos(\hat{\alpha})\cos(\hat{\gamma})-\sin(\hat{\alpha})\cos(\hat{\beta})\sin(\hat{\gamma}))^{2}\nonumber\\
 &\;\;\;\;+\text{\ensuremath{\chi}}_{3}\sin^{2}(\hat{\alpha})\sin^{2}(\hat{\beta})\bigg].
\end{alignat}
For $b_{x}=b_{y}$ we obtain circular polarization, while for $b_{x}=0$
or $b_{y}=0$ we obtain linear polarization along the $y$ or $x$
axis, respectively.

We now assume that the field $\bm{\hat{E}}_{d}$ is coherent and make the replacement $\hat{a}\rightarrow a$, where $a$ on the right hand-side
denotes a $\mathbb{C}$-value, which simplifies Eq.~(\ref{eq:gradient}) to the potential 
$\hat{H}_{\text{grad}}=-V\epsilon_{0}E_{0}^{2}\vert\hat{u}\vert^{2}\vert a\vert^{2}(\mathbf{\boldsymbol{\epsilon}}_{d}^{*})^{\top}\hat{\chi}\mathbf{\boldsymbol{\epsilon}}_{d}$.
In a more refined analysis one should also consider the effect of quantum fluctuations of the incoming field, i.e. $\hat{a}=a+\delta \hat{a}$,  where $\delta \hat{a}$ denote the quantum fluctuations. In particular, the $\delta \hat{a}$ contribution could lead to additional decoherence effects for the nanoparticle. We leave a more refined analysis, taking into account the quantum nature of the incoming optical field, for future research~\cite{greiner1997classical}. 

We now want to express the gradient potential in terms of experimentally controllable parameters. To this end suppose that the transverse cross-section of the beam is given by $\sigma_{L}$. In this case we
have that 
\begin{equation}
\epsilon_{0}E_{0}^{2}\vert a\vert^{2}=\frac{P}{c\sigma_{L}},\label{eq:power}
\end{equation}
where $P$ is the laser power, and $c$ is the speed of light. Using Eq.~\eqref{eq:power} we then immediately find
$\hat{H}_{\text{grad}}=-\frac{V P}{c\sigma_{L}}\vert\hat{u}\vert^{2}(\mathbf{\boldsymbol{\epsilon}}_{d}^{*})^{\top}\hat{\chi}\mathbf{\boldsymbol{\epsilon}}_{d}$.
 One can then consider a generic expansion
of $\vert u\vert^{2}$ up to a given order $\mathcal{O}((\vert \hat{\mathbf{r}} \vert/ l)^n)$:
\begin{equation}
\frac{VP}{c\sigma_{L}}\vert\hat{u}\vert^{2}=\sum_{k+l+m\leq n}c_{k,l,m}\hat{x}^{k}\hat{y}^{l}\hat{z}^{m},
\label{expansion}
\end{equation}
where $n\in\mathbb{N}$, and $l$ has dimensions of length. In general we have $\frac{(n+2)(n+1)}{2}$ free parameters up to and including order $n$. 

For example, one can consider a slightly modified Gaussian mode:
\begin{equation}
\hat{u}(\hat{r})=\frac{w_{0}}{w(\hat{z})}\text{exp}\left(-\frac{a_{1}\hat{x}^{2}+a_{2}\hat{y}^{2}}{w(\hat{z})^{2}}\right)e^{ik\hat{z}},\label{eq:umode2}
\end{equation}
where $a_{1}$, $a_{2}$ are two adimensional parameters that quantify the asymmetry, $w(\hat{z})=w_{0}\sqrt{1+\left(\frac{\hat{z}}{z_{R}}\right)^{2}}$,
$w_{0}$ is the beam waist, $k=\frac{2\pi}{\lambda}$, and $\lambda$ is the laser wavelength. In this case, assuming $w_0 \sim \lambda$ and $z_R \sim \lambda$, the relevant length scale for the expansion in Eq.~\eqref{expansion} is given b $l=\lambda$.  The asymmetry between $\hat{x}$ and $\hat{y}$ could arise for example due to the use of elliptical polarization~\cite{so2016tuning} or simply due to misalignment of the optical elements. In Eq.~\eqref{eq:umode2} we have for concreteness considered a travelling wave ($e^{ik\hat{z}}$), but an experimental situation with a standing wave can be described in a similar fashion. When the particle is confined close to the center of the trap, i.e. $ \frac{\vert\hat{x}\vert}{\lambda}$,$\frac{\vert\hat{y}\vert}{\lambda}$ and $ \frac{\vert\hat{z}\vert}{\lambda}$ are small, then only the harmonic terms are manifest in the dynamics of the nanoparticle ($c_{2,0,0}\propto-\frac{2a_1}{ w_{0}^{2}}$, $c_{0,2,0}\propto-\frac{2 a_2}{ w_{0}^{2}}$, and $c_{0,0,2}\propto-\frac{1}{z_{R}^{2}}$). On the other hand, if the nanoparticle starts exploring a larger region of the trap, then the first nonlinear terms start to become important, i.e. the quartic terms ($c_{4,0,0}\propto\frac{2 a_1^2}{ w_0^4}$, $c_{0,4,0}\propto\frac{2 a_2^2}{ w_0^4}$,  and $c_{0,0,4}\propto\frac{1}{z_{R}^{4}}$), and the cross coupling terms ($c_{2,2,0}\propto\frac{4 a_1 a_2}{w_{0}^{4}}$, $c_{2,0,2}\propto\frac{4a_1}{w_{0}^{2}z_{R}^{2}}$, and $c_{0,2,2}\propto\frac{4 a_2}{w_{0}^{2}z_{R}^{2}}$ ).

\subsubsection{Scattering terms\label{subsec:Scattering-terms}}

We consider the term $\propto\bm{\hat{E}}_{f}^{\top}\hat{\chi}\bm{\hat{E}}_{d}$,
where $\bm{\hat{E}}_{d}$ and $\bm{\hat{E}}_{f}$ are given in Eqs.~(\ref{eq:ed})
and (\ref{eq:ef}), respectively (as discussed below Eq.~\eqref{eq:ef} we consider the continuum limit, i.e. $V_{q}\rightarrow \infty$). This term, after tracing out the
free field degrees of freedom gives a decoherence term~\cite{agarwal2012quantum}.
Specifically, from Eqs.~(\ref{eq:coupling}), (\ref{eq:total_field})
we obtain the interaction Hamiltonian
\begin{equation}
\hat{H}_{\text{scattering}}=\sum_{\nu,\mu}\int \frac{d\bm{k}}{(2\pi)^3}\hat{B}_{\bm{k},\nu,\mu}\hat{S}_{\bm{k},\nu,\mu},\label{eq:scattering}
\end{equation}
where
\begin{equation}
\hat{B}_{\bm{k},\nu,\mu}=\begin{cases}
\hat{a}_{\bm{k},\nu}\,, & \text{for}\,\mu=0,\\
\hat{a}_{\bm{k},\nu}^{\dagger}\,, & \text{for}\,\mu=1,
\end{cases}
\end{equation}
are the bath operators, and 
\begin{equation}
\hat{S}_{\bm{k},\nu,\mu}=\begin{cases}
-i\sqrt{\frac{\hbar\omega_{k}}{2\epsilon_{0}}}e^{i\bm{k}\cdot\hat{\bm{r}}}\bm{\epsilon}_{\bm{k},\nu}^{\top}(\epsilon_{0}V\hat{\chi}\bm{\hat{E}}_{d})\,, & \text{for}\,\mu=0,\\
i\sqrt{\frac{\hbar\omega_{k}}{2\epsilon_{0}}}e^{-i\bm{k}\cdot\hat{\bm{r}}}\bm{\epsilon}_{\bm{k},\nu}^{*\top}(\epsilon_{0}V\hat{\chi}\bm{\hat{E}}_{d})\,, & \text{for}\,\mu=1,
\end{cases}
\end{equation}
are the system operators. We assume a zero temperature bath (corresponding to an initially empty bath):
\begin{alignat}{1}
\langle\hat{a}_{\bm{k},\nu},\hat{a}_{\bm{k}',\nu'}\rangle & =\langle\hat{a}_{\bm{k},\nu}^{\dagger},\hat{a}_{\bm{k}',\nu'}^{\dagger}\rangle=\langle\hat{a}_{\bm{k},\nu}^{\dagger},\hat{a}_{\bm{k}',\nu'}\rangle=0\,,\\
\langle\hat{a}_{\bm{k},\nu},\hat{a}_{\bm{k}',\nu'}^{\dagger}\rangle & =\delta^{(3)}(\bm{k}-\bm{k}')\delta_{\nu,\nu'}\,.\label{eq:commutator}
\end{alignat}
The assumption of zero bath temperature can be understood by noting that the bath is associated to the scattered photons: before the event of scattering of an incoming photon takes place, the bath consists of unpopulated modes, i.e. there are no scattered photons. Once a photon is then scattered, it populates a particular mode of the bath, but under the assumption of no self interaction between the bath modes, the bath for the next scattered photon immediately resets to an empty bath. Loosely speaking, one can think that two consecutive scattered photons are distant in time such that it is possible to account for them individually, at least as far as the overall effect on the nanoparticle's dynamics is concerned. Moreover, if the incoming field $\hat{\bm{E}}_d$, assumed classical, scatters into $\hat{\bm{E}}_d$, this do not lead to decoherence terms, but is already accounted for by the unitary gradient terms in Sec.~\ref{ssgt}.

In the Born Markov approximation, assuming the particle degrees of
freedom are not evolving during photon scattering (we assume that
the incoming and scattered wavelengths are the same, i.e. Rayleigh scattering),
making the rotating wave approximation (we time-average over the fast oscillations of the optical fields), supposing that the field
$\hat{\bm{E}}{}_{d}$ is coherent (we make the replacement $\hat{a}\rightarrow a$,
where $a$ on the right hand-side denotes a $\mathbb{C}$-value), using
Eq.~(\ref{eq:power}) and Eqs.~(\ref{eq:scattering})-(\ref{eq:commutator}),
we eventually obtain the Lindblad dissipator:
\begin{alignat}{1}
\mathcal{L}_{\text{scattering}}[\,\cdot\,]=&\gamma_{s}\sum_{\nu}\int d\bm{n}\bigg(  \hat{A}_{\bm{n},\nu}\,\cdot\,\hat{A}_{\bm{n},\nu}^{\dagger}\nonumber\\
&-\frac{1}{2}\left\{ \hat{A}_{\bm{n},\nu}^{\dagger}\hat{A}_{\bm{n},\nu},\,\cdot\,\right\} \bigg),\label{eq:scattering2}
\end{alignat}
where
\begin{equation}
\hat{A}_{\bm{n},\nu}=(\bm{\epsilon}_{k\bm{n},\nu}^{*\top}\hat{\chi}\bm{\epsilon}_{d})\hat{u}e^{i\bm{k}\cdot\hat{\bm{r}}},\label{eq:scatter_operator}
\end{equation}
and
\begin{equation}
\gamma_{s}=\frac{\tilde{\sigma}_{R}}{\sigma_{L}}\frac{P}{\hbar\omega_{L}}\label{eq:scattering_rate}
\end{equation}
is the scattering rate.  $\bm{n}$ denotes the unit
vector and $\tilde{\sigma}_{R}=\frac{\pi^{2}V_{0}^{2}}{\lambda^{4}}$
is an effective cross-section area.  For the case of an isotropic polarizable point
particle, Eq.~\eqref{eq:scattering2} reduces to the dissipator considered in~\cite{nimmrichter2014macroscopic,rashid2017wigner}:
in particular, we also re-obtain the Rayleigh cross-section $\sigma_{R}=\frac{24\pi^3V_0^2}{\lambda^4}\left(\frac{\epsilon_R-1}{\epsilon_R+2}\right)^2$, where $\epsilon_R$ is  the dielectric function, by combing the factors contained in $\tilde{\sigma}_{R}$ and $\chi$. The case of linear rotors with linearly polarized light, and the case
of arbitrary rotors with unpolarized light has been discussed in~\cite{stickler2016rotranslational,stickler2016spatio}
and~\cite{papendell2017quantum}, respectively.

\subsection{Gas collisions}
To account for the interaction with the gas of particles we suppose that the optically levitated particle is a many-body rigid system composed of $n$ particles. Specifically, we model the effect of gas collisions on this system using the dissipative Caldeira-Leggett master
equation~\cite{caldeira1985physica,breuer2002theory}:
\begin{alignat}{1}
&\mathcal{L}_{\text{collisional}}[\,\hat{\rho}\,]=  \frac{i\gamma_{c}}{2\hbar}\sum_{j=1}^{n}\left[\hat{\boldsymbol{r}}_{n}\cdot\hat{\boldsymbol{p}}_{n}+(\hat{\boldsymbol{r}}_{n}\cdot\hat{\boldsymbol{p}}_{n})^{\dagger},\hat{\rho}\right]
\nonumber\\
&\quad+\frac{4mk_{b}T\gamma_{c}}{\hbar^{2}}\sum_{j=1}^{n}\left(\hat{\tilde{\boldsymbol{L}}}_{n}\cdot\rho\hat{\tilde{\boldsymbol{L}}}_{n}^{\dagger}-\frac{1}{2}\left\{ \hat{\tilde{\boldsymbol{L}}}_{n}\cdot\hat{\tilde{\boldsymbol{L}}}_{n}^{\dagger},\hat{\rho}\right\} \right)\label{eq:cl}
\end{alignat}
where $\hat{\boldsymbol{r}}_{n}$ and $\hat{\boldsymbol{p}}_{n}$
are the position and momentum operators of particle $n$, respectively,
$m$ is the mass of a single particle, $\gamma_{c}$ is the collision
rate (assumed for simplicity the same for each particle), $k_{b}$
is Boltzman constant, $T$ is the temperature of the gas, and

\begin{equation}
\hat{\tilde{\boldsymbol{L}}}_{n}=\hat{\boldsymbol{r}}_{n}+\frac{i\hbar}{4mk_{b}T}\hat{\boldsymbol{p}}_{n}
\end{equation}
We now change to the center-of-mass (c.m.) coordinates:
\begin{alignat}{1}
\hat{\boldsymbol{r}}_{j} & =\hat{\boldsymbol{r}}+\hat{\tilde{\boldsymbol{r}}}_{j},\label{eq:rn}\\
\hat{\boldsymbol{p}}_{j} & =\frac{m}{M}\hat{\boldsymbol{p}}+\hat{\tilde{\boldsymbol{p}}}_{j},\label{eq:pn}
\end{alignat}
where $\hat{\boldsymbol{r}}$, $\hat{\boldsymbol{p}}$, $\hat{\tilde{\boldsymbol{r}}}_{n}$, $\hat{\tilde{\boldsymbol{p}}}_{n}$
are the c.m. position, c.m. momentum, relative position of $n$-th
particle, relative momentum of $n$-th particle, operators, respectively,
and $M=nm$ is the total mass. We now use Eqs.~(\ref{eq:rn}), (\ref{eq:pn}),
and the relations $\sum_{j=1}^{n}\hat{\tilde{\boldsymbol{r}}}_{j}=0$,
$\sum_{j=1}^{n}\hat{\tilde{\boldsymbol{p}}}_{j}=0$, to decouple c.m. and relative degrees of freedom in Eq.~(\ref{eq:cl}):
\begin{equation}
\mathcal{L}_{\text{collisional}}[\,\cdot\,]=\mathcal{L}_{\text{collisional}}^{(t)}[\,\cdot\,]+\mathcal{L}_{\text{collisional}}^{(r)}[\,\cdot\,],\label{eq:cl2}
\end{equation}
where $\mathcal{L}_{\text{collisional}}^{(t)}[\,\cdot\,]$ and $\mathcal{L}_{\text{collisional}}^{(r)}[\,\cdot\,]$
denote the dissipator on translations and, as discussed below, rotations,
respectively. Specifically, we find the following dissipator for translations:
\begin{alignat}{1}
&\mathcal{L}_{\text{collisional}}^{(t)}[\,\hat{\rho}\,]=  \frac{i\gamma_{c}}{2\hbar}\left[\hat{\boldsymbol{r}}\cdot\hat{\boldsymbol{p}}+(\hat{\boldsymbol{r}}\cdot\hat{\boldsymbol{p}})^{\dagger},\hat{\rho}\right] \nonumber\\
&\quad+\frac{4Mk_{b}T}{\hbar^{2}}\gamma_{c}\left(\hat{\tilde{\boldsymbol{L}}}\cdot\rho\hat{\tilde{\boldsymbol{L}}}^{\dagger}-\frac{1}{2}\left\{ \hat{\tilde{\boldsymbol{L}}}\cdot\hat{\tilde{\boldsymbol{L}}}^{\dagger},\hat{\rho}\right\} \right),\label{eq:clt}
\end{alignat}
where $\hat{\tilde{\boldsymbol{L}}}=\hat{\boldsymbol{r}}+\frac{i\hbar}{4Mk_{b}T}\hat{\boldsymbol{p}}$.
Under the assumption
of a rigid body we eventually find the following dissipator for rotations:
\begin{alignat}{1}
\mathcal{L}_{\text{collisional}}^{(r)}[\,\hat{\rho}\,]=&\frac{4mk_{b}T}{\hbar^{2}}\gamma_{c}\sum_{\zeta=1}^{3}\tilde{D}_{\zeta}\bigg(\left[\hat{\tilde{\boldsymbol{C}}}_{\zeta}\cdot\hat{\rho}\hat{\tilde{\boldsymbol{C}}}_{\zeta}^{\dagger}\right]
\nonumber\\
&-\frac{1}{2}\left\{ \hat{\tilde{\boldsymbol{C}}}_{\zeta}^{\dagger}\cdot\hat{\tilde{\boldsymbol{C}}}_{\zeta},\hat{\rho}\right\} \bigg),\label{eq:clr}
\end{alignat}
where 
\begin{equation}
\hat{\tilde{\boldsymbol{C}}}_{\zeta}=\hat{F}\mathbf{e}_{\zeta}-\frac{i\hbar}{4k_{b}T}\hat{F}L_{\zeta}I^{-1}\hat{F}^{\top}(\hat{N}^{\top})^{-1}\hat{\bm{\pi}},\label{eq:azeta}
\end{equation}
$\boldsymbol{e}_{\zeta}$ is the unit vector along the $\zeta$-axis, $L_{\zeta}$
is the generator of rotations about the $\zeta$-axis, and 
\begin{equation}
\tilde{D}_{\zeta}=(\frac{1}{2}\text{tr}I-I_{\zeta}).
\end{equation}
The moment of inertia tensor $I$, the Euler parametrization $\hat{F}$ of a generic rotation, and the matrix $\hat{N}$ have been defined in Sec.~\ref{sec:fh}.
For later convenience, we also define the operators:
\begin{alignat}{1}
\hat{L}_{j} & =\frac{i\sqrt{4Mk_{b}T}}{\hbar}\hat{\tilde{\boldsymbol{L}}}\cdot\boldsymbol{e}_{j},\\
\hat{\boldsymbol{C}}_{\zeta,j} & =\frac{i\sqrt{4k_{b}T\tilde{D}_{\zeta}}}{\hbar}\hat{\tilde{\boldsymbol{C}}}_{\zeta}\cdot\boldsymbol{e}_{j}.
\end{alignat}
The case of rotational diffusion without friction is discussed in~\cite{papendell2017quantum}, while the dissipator in Eq.~\eqref{eq:clr} has been derived in \cite{stickler2017rotational}.

\subsection{Non-inertial terms}

For completeness we also include the non-inertial term, which arises
in Earth-bound laboratories. Specifically, we consider the following
contribution to the Hamiltonian:
\begin{equation}
\hat{H}_{\text{ni}}=Mg\hat{x},\label{eq:ni}
\end{equation}
where $M$ is the total mass, and $g$ is the gravitational acceleration.
Although the contribution from this term is typically much smaller
than from light-matter interactions and gas collisions, it can become relevant in
certain experimental settings~\cite{kuhn2015cavity,hebestreit2017sensing,stickler2018orientational}.

\section{Detection for ro-translation} \label{sec:Dynamics}

In this section we combine the terms from the previous Sec.~\ref{sec:description-of-the} and discuss the resulting dynamics. In particular, we consider the unconditional dynamics, i.e. without a detector keeping track of the intensity gathered from the collected scattered photons, and the dynamics
conditioned upon the measured intensity in a general dyne detection.
We then apply the obtained formulae to construct to a toy model of homodyne detection.

\subsection{Dyne detection}\label{subsec:Dyne-detection}

The dynamics of the optically levitated particle is given
by:
\begin{alignat}{1}
\dot{\hat{\rho}}=&-\frac{i}{\hbar}[\hat{H}_{\text{free}}+\hat{H}_{\text{gradient}}+\hat{H}_{\text{ni}},\hat{\rho}]\nonumber\\
&+\mathcal{L}_{\text{scattering}}[\hat{\rho}]+\mathcal{L}_{\text{collisional}}[\hat{\rho}]\label{eq:master}
\end{alignat}
where $\hat{H}_{\text{free}}$, $\hat{H}_{\text{grad}}$, $\mathcal{L}_{\text{scattering}}[\,\cdot\,]$,
and $\mathcal{L}_{\text{collisional}}[\,\cdot\,]$ are defined in
Eqs.~(\ref{eq:free}), (\ref{eq:gradient}), (\ref{eq:scattering2}),
and (\ref{eq:cl2}), respectively, and $\hat{H}_{\text{ni}}$ is given in Eq.~\eqref{eq:ni}. We will refer to Eq.~(\ref{eq:free})
as the unconditional dynamics, and to the state $\hat{\rho}$ as the unconditional state.

However, usually one collects part of the scattered light to update the knowledge about the state of the system. Here we consider the case when the scattered light interfers with a classical local oscillator before detection, namely, dyne detection (see Fig.~\ref{fig:setup}(c)). A simple example of this type of approach is given by homodyne detection~\cite{rashid2017wigner}.

The detected photo-current (signal) allows to continuously update the description of the system: we will refer to the resulting state $\hat{\rho}_{c}$ as the conditional state.  Mathematically we can describe this by considering an unraveling of the photon scattering term $\mathcal{L}_{\text{scattering}}[\hat{\rho}]$ in Eq.~(\ref{eq:master}). The most general diffusive unraveling,
also known as the Belavkin  equation, is given by (in It\^{o} form)~\cite{wiseman2001complete,WISEMAN2001227}:
\begin{alignat}{1}
d\hat{\rho}_{c}=&\gamma_{s}\sum_{\nu=1}^{2}\int d\bm{n}\mathcal{D}[\hat{A}_{\bm{n},\nu}]\hat{\rho}_{c}dt\nonumber\\
&+\sqrt{\gamma_{s}}\sum_{\nu=1}^{2}\int d\bm{n}\mathcal{H}[\hat{A}_{\bm{n},\nu}dW_{\bm{n},\nu}^*]\hat{\rho}_{c},\label{eq:unraveling}
\end{alignat}
where~\cite{PhysRevA.47.1652}
\begin{alignat}{1}
\mathcal{D}[\hat{K}]\,\cdot\, & =\hat{K}\,\cdot\,\hat{K}^{\dagger}-\frac{1}{2}\left\{ \hat{K}^{\dagger}\hat{K},\,\cdot\,\right\} ,\label{eq:d}\\
\mathcal{H}[\hat{K}]\,\cdot\, & =\hat{K}\,\cdot\,+\,\cdot\,\hat{K}^{\dagger}-\text{tr}[\hat{K}\,\cdot\,+\,\cdot\,\hat{K}^{\dagger}]\,\cdot\,,\label{eq:h}
\end{alignat}
and $\hat{K}$ denotes an operator. Note that the first term on the
right hand-side of Eq.~(\ref{eq:unraveling}) corresponds to $\mathcal{L}_{\text{scattering}}[\hat{\rho}]$
. $W_{\bm{n},\nu}$ are $\mathbb{C}$-valued, zero mean Wiener processes
with correlations:
\begin{alignat}{1}
\mathbb{E}[dW_{\bm{n},\nu}dW_{\bm{n}',\nu'}^{*}] & =dt\eta_{(\bm{n},\nu),(\bm{n}',\nu')},\label{eq:dwdws}\\
\mathbb{E}[dW_{\bm{n},\nu}dW_{\bm{n}',\nu'}] & =dt\varXi_{(\bm{n},\nu),(\bm{n}',\nu')}\,,\label{eq:dwdw}
\end{alignat}
where the only non-zero elements of $\eta$ are $\eta_{(\bm{n},\nu),(\bm{n},\nu)}\in[0,1]$,
$\varXi$ has $\mathbb{C}$-valued entries, $\varXi_{(\bm{n},\nu),(\bm{n}',\nu')}=\varXi_{(\bm{n}',\nu'),(\bm{n},\nu)}$,
and 
\begin{equation}
\frac{1}{2}\left(\begin{array}{cc}
\eta+\mathbb{\text{Re}}(\varXi) & \mathbb{\text{Im}}(\varXi)\\
\mathbb{\text{Im}}(\varXi) & \eta-\mathbb{\text{Re}}(\varXi)
\end{array}\right)
\end{equation}
is positive semi-definite. The photo-currents associated to Eq.~(\ref{eq:unraveling})
are given by:
\begin{alignat}{1}
J_{\bm{n},\nu}dt=&\text{tr}\bigg[\sum_{\nu'=1}^{2}\int d\bm{n}'\bigg(\eta_{(\bm{n},\nu),(\bm{n}',\nu')}\hat{A}_{\bm{n}',\nu'}\nonumber\\
&+\varXi_{(\bm{n},\nu),(\bm{n}',\nu')}\hat{A}_{\bm{n}',\nu'}^{\dagger}\bigg)\hat{\rho}_{c}\bigg]dt+dW_{\bm{n},\nu}.\label{eq:currents}
\end{alignat}

Eqs.~\eqref{eq:unraveling} and \eqref{eq:currents} is the conventional way of presenting the conditional dynamics: the stochastic nature of the dynamics and of the photo-current is explicit, where the stochasticity is due to the weak (imprecise) measurements of the system. However, one can also combine Eqs.~\eqref{eq:unraveling} and \eqref{eq:currents} in a single equation that explicitly shows the dependency of the conditional dynamics on the measured photo-currents $J_{\bm{n},\nu}$. In particular, one can invert Eq.~\eqref{eq:currents} to obtain the expression of $dW_{\bm{n},\nu}$ as a function of the measured photo-currents $J_{\bm{n},\nu}$, i.e. $dW_{\bm{n},\nu}(J_{\bm{n},\nu})$, which can be used to eliminate the Wiener processes $dW^*_{\bm{n},\nu}$ from Eq.~(\ref{eq:unraveling}):
\begin{alignat}{1}
d\hat{\rho}_{c} =&\gamma_{s}\sum_{\nu=1}^{2}\int d\bm{n}\mathcal{D}[\hat{A}_{\bm{n},\nu}]\hat{\rho}_{c}dt\nonumber\\
&+\sqrt{\gamma_{s}}\sum_{\nu=1}^{2}\int d\bm{n}\mathcal{H}[\hat{A}_{\bm{n},\nu}dW_{\bm{n},\nu}(J_{\bm{n},\nu})^*]\hat{\rho}_{c},\label{eq:unraveling2}
\end{alignat}
The evolution of the conditional state $\hat{\rho}_{c}$ in Eq.~\eqref{eq:unraveling2} now explicitly depends on the currents $J_{\bm{n},\nu}$, which are inputs of the equation of motion. The conditional dynamics in Eq.~\eqref{eq:unraveling2} can be readily used for tracking or simulating the conditional state of the system~\cite{setter2017real,ralph2017dynamical}.

The full conditional dynamics can be obtained by adding the Hamiltonian terms ($\hat{H}_{\text{free}}$, $\hat{H}_{\text{grad}}$ and $\hat{H}_{\text{ni}}$) and the non-unitary contribution from gas collisions ($\mathcal{L}_{\text{collisional}}$) to the right hand-side of Eqs.~(\ref{eq:unraveling}) or \eqref{eq:unraveling2}. Discontinuous unravellings, where each photon triggers a discontinuous update of the conditional state, could be treated in a similar way. 

In general, the currents $J_{\bm{n},\nu}$ are $\mathbb{C}$-valued
and thus cannot be directly associated to the intensity current measured
by a physical detector: these can be reconstructed from the $\mathbb{R}$-valued
currents $\text{Re}(J_{\bm{n},\nu})$ and $\text{Im}(J_{\bm{n},\nu})$, e.g. see heterodyne detection in~\cite{wiseman2009quantum}.
In the next section we consider the case of homodyne detection,
which is a special case of the formalism used in this section, where
we obtain explicit expression for the physical photo-currents. 

\subsection{Homodyne detection model}\label{sec:dyne}

In order to discuss a detection model we have to specify the measuring operator(s). In general, the measuring operator will be a functional of the system degrees
of freedom as well as of the experimental setting, i.e. $\mathcal{A}[\hat{\boldsymbol{r}},\hat{\boldsymbol{\phi}};\text{exp.setting}]$.
For example, only some of the scattered photons are collected by optical
elements: these are then recorded by a physical detector, where the
detector's efficiency, orientation, distance, size, and integration
time, all affect the measured signal. Here we consider a simplified
detector model, completely characterized by the operator $\sqrt{\eta\gamma_s}\sum_{\nu=1}^{2}\int_{S}d\bm{n}\hat{A}_{\bm{n},\nu},$
where $S$ denotes the surface of a toy detector, $\gamma_s$ is defined in Eq.~\eqref{eq:scattering_rate}, and $\eta$ is the detector's efficiency, i.e. we are considering the case when the efficiency matrix $\eta$ introduced in Sec.~\ref{subsec:Dyne-detection} is proportional to the identity matrix, and completely characterized by a single number, which we also label as $\eta \in [0,1]$ (see Fig.~\ref{fig:setup}(a)). In this case, as we show below, the total photo-current is of the form $\sum_{\nu}\int_{S}d\bm{n}\,J_{\boldsymbol{n},\nu}$, where $J_{\boldsymbol{n},\nu}$ is associated to $\hat{A}_{\bm{n},\nu}$.

This total photo-current, which we label as $J$, can be considered as a toy model for the experimental configuration in~\cite{rashid2017wigner}. Loosely speaking, optical elements, such as a paraboloidal mirror, collect the scattered photons and direct 
them towards the beam splitter: this conceals, at least partially,
the information about the scattering direction $\boldsymbol{n}$ and
polarization $\nu$. We denote the annihilation operator for the corresponding collective mode by $\hat{a}_{\text{out}}$, i.e. the annihilation operator of all the photons travelling towards the detector. At the beam splitter the signal from the scattered
photons is combined with the local oscillator $a_{\text{LO}}$ (a $\mathbb{C}$-value) from which we obtain the current $J$ (see Fig.~\ref{fig:setup}(c)). Here we are supposing that the local oscillators $\left(a_{\bm{n},\nu}\right)_{\text{LO}}$, for each  direction $\bm{n}$ and polarization $\nu$, can be approximated by a single local oscillator $a_{\text{LO}}$. To obtain a more refined model of detection in this specific experimental situation, or to adapt it to describe a different experimental setup, one would need to take into account the specific details of the experiment and repeat the analysis, e.g. by imposing the specific boundary conditions. 

We can now apply the general procedure discussed in the previous Sec.~\ref{subsec:Dyne-detection}.
Specifically, for each dissipator term $D[\hat{A}_{\bm{n},\nu}]$
we have to consider the corresponding noise term $\mathcal{H}[\hat{A}_{\bm{n},\nu}dW_{\bm{n},\nu}]$,
where we assume that $W_{\bm{n},\nu}$ are $\mathbb{R}$-valued and
independent, since they are associated to different modes. As already mentioned above, we also suppose that each mode is detected with the same efficiency $\eta\in[0,1]$, which simplifies Eqs.~(\ref{eq:dwdws}) and (\ref{eq:dwdw}) to 
\begin{equation}
\mathbb{E}[dW_{\bm{n},\nu}dW_{\bm{n}',\nu'}]=\eta dt\delta_{\nu,\nu'}\delta^{(2)}(\bm{n}-\bm{n}').
\end{equation}
It is then straightforward to obtain the equation for the conditional
state (in It\^{o} form):
\begin{alignat}{1}
d\hat{\rho}_{c}=&  -\frac{i}{\hbar}[\hat{H}_{\text{free}}+\hat{H}_{\text{gradient}}+\hat{H}_{\text{ni}},\hat{\rho}_{c}]dt\nonumber \\
 &  +\gamma_{c}\sum_{j=1}^{3}\mathcal{D}[\hat{L}_{j}]\hat{\rho}_{c}dt+\gamma_{c}\sum_{\zeta,j=1}^{3}\mathcal{D}[\hat{\boldsymbol{C}}_{\zeta,j}]\hat{\rho}_{c}dt \nonumber\\
 & +\gamma_{s}\sum_{\nu=1}^{2}\int d\bm{n}\mathcal{D}[\hat{A}_{\bm{n},\nu}]\hat{\rho}_{c}dt \nonumber\\
&+\sqrt{\gamma_{s}}\mathcal{H}[\sum_{\nu=1}^{2}\int_{S}d\bm{n}\hat{A}_{\bm{n},\nu}]\hat{\rho}_{c}dW
.\label{eq:sme}
\end{alignat}
$W$ is a zero mean, $\mathbb{R}$-valued Wiener process with correlation 
\begin{equation}
\mathbb{E}[dWdW]  =2\Omega \eta dt,
\label{meganoise}
\end{equation}
where $\Omega=\int_{S}d\bm{n}$, and the factor $2$ reflects the
fact that both independent polarizations are detected. Using Eq.~\eqref{eq:currents}, summing all the currents, we finally obtain that the  state $\hat{\rho}_{c}$ in Eq.~(\ref{eq:sme}) is conditioned on the following photo-current:
\begin{equation}
J dt=\eta\sqrt{\gamma_{s}}\text{Tr}\left[\sum_{\nu=1}^{2}\int_{S}d\bm{n}\left(\hat{A}_{\bm{n},\nu}+\hat{A}_{\bm{n},\nu}^{\dagger}\right)\hat{\rho}_{c}\right]dt+dW.\label{eq:current}
\end{equation}
We recover Eq.~(\ref{eq:master}) from Eq.~(\ref{eq:sme}) by taking the expectation value $\mathbb{E}[\,\cdot\,]$ over the
noise realizations. In case $dW$ is obtained from $J$ by inverting Eq.~\eqref{eq:current} one has to repeat the experiment or simulation to build enough statistics for J in order to recover Eq.~(\ref{eq:master}). 

\subsubsection{Heisenberg picture}
The above derivation in the Schr\"{o}dinger picture, on the one hand, has the advantage that it clearly shows the effect of photon detection on the nanoparticle, i.e. one inverts Eq.~\eqref{eq:current} and then inserts the expression for $dW(J)$ in Eq.~\eqref{eq:sme},
on the other hand, it does not provide an intuitive picture of the interaction between the photons and the nanoparticle. This becomes more apparent in Heisenberg picture using the input-output formalism~\cite{gardiner1985input,gardiner2004quantum,
wiseman2009quantum}. In a nusthell, an incoming photon $\hat{a}$, associated to the field $\hat{\bm{E}}_d$ interacts with the nanoparticle, which generates a signature in the scattered photon $\hat{a}_{\bm{n},\nu}$ associated to the field $\hat{\bm{E}}_f$. In particular, one labels the operator of the scattered photon, before and after the event of scattering takes place, as the input operator $\left(\hat{a}_{\bm{n},\nu}\right)_{\text{in}}$ and output operator $\left(\hat{a}_{\bm{n},\nu}\right)_{\text{out}}$, respectively.   As the particle scatters the incoming photon, the input operator transforms to the output operator according to the following relation:
\begin{equation}
\left(\hat{a}_{\bm{n},\nu}\right)_{\text{out}}=\left(\hat{a}_{\bm{n},\nu}\right)_{\text{in}}+\sqrt{\gamma_{s}}\hat{A}_{\bm{n},\nu}\,,\label{eq:inout}
\end{equation}

The modelling of inefficient detection is slightly more involved in the Heisenberg picture. To show the close analogy with the Schr\"{o}dinger picture analysis it is convenient to define the input quantum noise operator $d\hat{a}_\text{in}=\sum_{\nu=1}^{2}\int d\bm{n}\left(\hat{a}_{\bm{n},\nu}\right)_{\text{in}} dt$, where we have  $[d\hat{a}_\text{in},d\hat{a}_\text{in}^\dagger]=2\Omega dt$, and to introduce a second auxiliary quantum noise operator $d\hat{v}$, such that $[d\hat{v},d\hat{v}^\dagger]=2\Omega dt$. Here we assume that the quantum noise operators act on the vacuum state of their corresponding bath. Loosely speaking we can think of $d\hat{a}_\text{in}$ as the quantum noise in case of a completely efficient detection, i.e. $\eta = 1$, which starts to become completely dominated by the noise $d\hat{v}$ at low efficiencies, i.e. $\eta \ll 1$. This can be seen mathematically by formally introducing a new quantum noise operator $d\hat{w}$:
\begin{alignat}{1}
d\hat{w}=\eta
\left(d\hat{a}_\text{in}
+d\hat{a}_\text{in}^\dagger \right)
+\sqrt{\eta(1-\eta)}\left(d\hat{v}+d\hat{v}^\dagger\right),
\end{alignat}
such that $\langle d\hat{w} \rangle=0$ and $\langle d\hat{w} d\hat{w}\rangle=2\Omega \eta dt$. The statistics of the photo-current $J$ in Eq.~(\ref{eq:current}) can then be recovered by considering its Heisenberg picture equivalent (see Fig.~\ref{fig:setup}(c)):
\begin{alignat}{1}
\hat{J}dt=&\eta\sqrt{\gamma_{s}} \sum_{\nu}\int_{S}d\bm{n}
 \left((\hat{A}_{\bm{n},\nu}+\hat{A}_{\bm{n},\nu}^\dagger \right)dt+d\hat{w}.
\label{eq:current2}
\end{alignat}
In particular, one can readily show that
\begin{equation}
\mathbb{E}[J]=\langle \hat{J}\rangle, \qquad 
\mathbb{E}[(J-\mathbb{E}[J])^2]=\langle (\hat{J}-\langle \hat{J}\rangle)^2 \rangle,
\label{JeJq}
\end{equation} 
where $\mathbb{E}[\;\cdot\;]$ denotes the stochastic expectation value with respect to different noise realizations, and
$\langle\;\cdot\;\rangle$ denotes the quantum trace operation with respect to the nanoparticle state and the vacuum states of the two baths. For more details see~\cite{gardiner2004quantum,wiseman2009quantum,jacobs2014quantum}.

\subsubsection{Classical currents}
It is useful to derive approximate photo-currents for a classical nanoparticle, e.g. for force and torque sensing applications. To this end we replace quantum observables $\hat{O}$ by their corresponding classical observables $O^{\text{(cl)}}$, and
the commutators with Poisson Brackets, i.e $[\,\cdot\,,\,\cdot\,]\rightarrow i\hbar\{\,\cdot\,,\,\cdot\,\}_{\text{Pb}}$. In particular, following this procedure, we obtain from Eq.~\eqref{eq:current}:
\begin{equation}
Jdt=2\eta\sqrt{\gamma_{s}}\sum_{\nu=1}^{2}\int_{S}d\bm{n}\text{Re}\left(A_{\bm{n},\nu}^{\text{(cl)}} e^{i\Delta \Phi} \right)dt+dW.\label{eq:current_cl}
\end{equation}
where we have introduced the phase $\Delta\Phi$ of the local oscillator. From Eq.~\eqref{eq:scatter_operator} we also readily obtain the classical scattering observable:
\begin{equation}
A_{\bm{n},\nu}^{(cl)}(\bm{r},\bm{\phi})=[\bm{\epsilon}_{k\bm{n},\nu}^{*\top}F(\bm{\phi})\chi F(\bm{\phi})^\top\bm{\epsilon}_{d}]u(\bm{r})e^{i\bm{k}\cdot\bm{r}}.\label{eq:scatter_operator_cl}
\end{equation} 

Let us now consider separately the position and angle depended factors in $A_{\bm{n},\nu}^{\text{(cl)}}$. We assume the modified Gaussian mode in Eq.~\eqref{eq:umode2} and suppose that $\frac{\vert\boldsymbol{r}\vert}{\lambda}$  is small. In particular, we consider the expansion up to order 
$\mathcal{O}(\vert\boldsymbol{r}\vert^{2})$:
\begin{alignat}{1}
u(\bm{r})e^{i\bm{k}\cdot\bm{r}}
\approx & 1 + i (k \bm{n} \bm{r}+k z) 
-k^2 {\bm n}\cdot {\bm r} z	
-\frac{1}{2} k^2 (\bm{n}\cdot \bm{r})^2
\nonumber\\
&-\frac{a_1 }{w_0^2}x^2-\frac{a_2 }{w_0^2}y^2
-\frac{z_R^2 k^2+2}{2 z_R^2} z^2 
\label{position}
\end{alignat}
We also decompose the susceptibility tensor $\chi$ (in the body frame) in the following form:
\begin{equation}
\chi=\chi_{0}(\mathcal{I}+\Delta\chi)
\label{orientation}
\end{equation}
where $\chi_{0}$ is the susceptibility in the limit of an isotropic particle, $\Delta\chi$ quantifies the degree of anisotropy, and $\mathcal{I}$ denotes the $3\times3$ identity matrix.
Using Eqs.~\eqref{eq:scatter_operator_cl}-\eqref{orientation} we can then decompose the expectation value of the photo-current in Eq.~\eqref{eq:current_cl} in four parts:
\begin{equation}
\mathbb{E}[J(\bm{r},\bm{\phi}; \Delta\Phi)]=J_{\text{0}}+J_{\text{T}}(\bm{r})+J_{\text{R}}(\bm{\phi})+J_{\text{RT}}(\bm{r},\bm{\phi}),
\label{current_cl_e}
\end{equation}
where $J_{\text{0}}$, $J_{\text{T}}$, $J_{\text{R}}$, and $J_{\text{RT}}$ denote a constant, a purely translational, a purely rotational, and the mixed ro-translational expectation values of the currents, respectively.

We first discuss the limit of an isotropic particle ($\Delta \chi \rightarrow 0$) such that the only non-trivial term in Eq.~\eqref{current_cl_e} is given by:
\begin{alignat}{1}
J_{\text{T}}(\bm{r};\Delta \Phi)  =&2\eta\chi_{0}\sqrt{\gamma_{s}}\int_{S}d\bm{n}\text{Re}\bigg[\sum_{\nu=1}^{2}\bm{\epsilon}_{k\bm{n},\nu}^{*\top}\bm{\epsilon}_{d} e^{i\Delta \Phi}\nonumber\\ 
&\big( i k( \bm{n} \bm{r}+ z) -k^2 {\bm n}\cdot {\bm r} z	 -\frac{1}{2} k^2 (\bm{n}\cdot \bm{r})^2\nonumber\\ 
&-\frac{a_1 }{w_0^2}x^2-\frac{a_2 }{w_0^2}y^2
-\frac{z_R^2 k^2+2}{2 z_R^2} z^2 \big)\bigg].\label{eq:JT}
\end{alignat}
In case of linearly polarized light $\bm{\epsilon}_{d}$ has $\mathbb{R}$-valued components and thus also $\sum_{\nu=1}^{2}\bm{\epsilon}_{k\bm{n},\nu}^{*\top}\bm{\epsilon}_{d}$ becomes $\mathbb{R}$-valued. By controlling the phase $\Delta\Phi$ of the local oscillator we can then decide to detect the position of the particle, i.e. the first term  ($\propto \bm{n} \bm{r}+ z$) on the second line of Eq.~\eqref{eq:JT}, or the squared value of position and cross-coupling terms, i.e. the last two terms on the second line and the last line of Eq.~\eqref{eq:JT}.

We next discuss the limit of small position oscillations ($\vert \bm{r}\vert \rightarrow 0$) such that the only non-trivial term in Eq.~\eqref{current_cl_e} is given by:
\begin{alignat}{1}
J_{\text{R}}({\bm \phi};\Delta \Phi)  = &2\eta\chi_{0}\sqrt{\gamma_{s}}\int_{S}d\bm{n}\text{Re}\bigg[\nonumber\\
&\sum_{\nu=1}^{2}\bm{\epsilon}_{k\bm{n},\nu}^{*\top}
F(\mathbf{\mathbf{\mathbf{\boldsymbol{\phi}}}})\Delta\chi F(\mathbf{\mathbf{\mathbf{\boldsymbol{\phi}}}})^{\top}\bm{\epsilon}_{d}
e^{i\Delta \Phi}\bigg].\label{eq:JR}
\end{alignat}
If we consider again linearly polarized light, i.e. $\sum_{\nu=1}^{2}\bm{\epsilon}_{k\bm{n},\nu}^{*\top}\bm{\epsilon}_{d}$ is $\mathbb{R}$-valued, then we 
see that $\Delta \Phi$ controls the amplitude of the photocurrent $J_{\text{R}}$, but not the measured observable. This is in different from the translational current $J_{\text{T}}$ in Eq.~\eqref{eq:JT}, where the phase $\Delta \Phi$ of the local oscillator controls the amplitude as well as the measured observable.

The correction current $J_{\text{RT}}({\bm r},{\bm \phi};\Delta \Phi)$ can be obtained by combing $J_{\text{T}}({\bm r};\Delta \Phi)$ together with $J_{\text{R}}({\bm \phi};\Delta \Phi)$: specifically, $J_{\text{RT}}$ can be formally obtained by inserting the terms on the second and third lines of Eqs.~\eqref{eq:JT} inside the square brackets of Eq.~\eqref{eq:JR}.

The formulae in Eqs.~\eqref{eq:JT} and \eqref{eq:JR} can be used for investigating the conversion between the measured homodyne current $J$ and the nanoparticle position ($\bm{r}$) and orientation ($\bm{\phi}$). To include explicitly the amplitude of the local oscillator one can follow the approach taken in~\cite{rashid2017wigner}, which can be readily extended to include ro-translations. Moreover, while in this section we have discussed currents based on the measurement of classical observables, the same analysis can be applied also for the current based on quantum observables of the quantum model discussed in the previous sections. In particular, one obtains an analogous separation of the currents in translational, rotational, and ro-translational terms, as discussed below Eq.~\eqref{current_cl_e}.

\section{Summary}\label{sec:Conclusion}

We have discussed the motion and detection of optically levitated
nanoparticles. Specifically, we have considered an anisotropic particle trapped
in an elliptically polarized Gaussian beam, and immersed in a bath
of gas particles. We have first introduced the dynamics of such systems using notions of classical electromagnetism and mechanics: the resulting ro-translational motion is driven (photon scattering), damped (gas particle collisions), as well as diffusive (photon scattering and gas particle collisions). We have then derived the complete quantum dynamics and discussed in detail the detection of the nanoparticle. Specifically, under the Born-Markov assumption we have obtained
the unconditional dynamics and the dynamics conditioned upon a general
dyne measurement. We have discussed the relation between the photo-currents, the measuring operators, and the dynamics both in the Schr\"{o}dinger, as well as in the Heisenberg picture. We have illustrated the use of the general formulae by constructing a toy model of homodyne detection. We have obtained approximate formulae, which could be used to extract the nanoparticle position and orientation from the measured signal. 

\begin{acknowledgments}
We also like to thank A. Bassi, M. Carlesso, and G. Gasbarri for discussions. We wish to thank for research funding, The Leverhulme Trust and the Foundational Questions Institute
(FQXi). This project has received funding from the European Union's Horizon 2020 research and innovation programme under grant agreement No 766900. We also acknowledge support by the EU COST action QTSpace (CA15220).
\end{acknowledgments}

\appendix

\section{Polarization of scattered light\label{sec:Polarization-of-scattered}}

In this section we briefly discuss the decomposition in Eq.~(\ref{eq:ef}).
Consider a fixed scattering direction $\mathbf{n}$ and the orthogonal
plane described by the tensor $\sum_{\nu}\bm{\epsilon}_{\bm{n},\nu}\otimes\bm{\epsilon}_{\bm{n},\nu}^{*}$,
i.e. the completeness relation in Eq.~(\ref{eq:cr}). We consider
two orthogonal axis in this plane, which we denote by $x$ and $y$,
and the corresponding unit vectors along these axis, which we denote
by $\boldsymbol{e}_{x}$ and $\boldsymbol{e}_{y}$, respectively.
Moreover, we require that $\boldsymbol{e}_{x}$, $\boldsymbol{e}_{y}$
and $\mathbf{n}$ form the directions of a right-handed coordinate
system.

In this coordinate system we can consider different decompositions.
Particularly simple is the linear decomposition:

\begin{equation}
\sum_{\nu}\bm{\epsilon}_{\bm{\mathbf{n}},\nu}\hat{a}_{\bm{k},\nu}=\boldsymbol{e}_{x}\hat{a}_{\bm{k},x}+\boldsymbol{e}_{y}\hat{a}_{\bm{k},y},\label{eq:linear}
\end{equation}
where $\hat{a}_{\bm{k},x}$, $\hat{a}_{\bm{k},y}$ denote annihilation
operators for photons with polarizations along $x$ and $y$, respectively.
Alternatively, we can consider the circular decomposition:

\begin{equation}
\sum_{\nu}\bm{\epsilon}_{\bm{\mathbf{n}},\nu}\hat{a}_{\bm{k},\nu}=\frac{1}{\sqrt{2}}\left(\begin{array}{c}
1\\
i\\
0
\end{array}\right)\hat{a}_{\bm{k},R}+\frac{1}{\sqrt{2}}\left(\begin{array}{c}
1\\
-i\\
0
\end{array}\right)\hat{a}_{\bm{k},L},\label{eq:circular}
\end{equation}
where $\hat{a}_{\bm{k},L}$, $\hat{a}_{\bm{k},R}$ denote annihilation
operators for left and right photons, respectively. Comparing the
two expressions in Eqs.~(\ref{eq:linear}) and (\ref{eq:circular})
we find:

\begin{alignat}{1}
\hat{a}_{\bm{k},L} & =\frac{\hat{a}_{\bm{k},x}+i\hat{a}_{\bm{k},y}}{\sqrt{2}},\\
\hat{a}_{\bm{k},R} & =\frac{\hat{a}_{\bm{k},x}-i\hat{a}_{\bm{k},y}}{\sqrt{2}}.
\end{alignat}
Similarly, one could also consider other decompositions, such as the
elliptical, and find the decomposition of corresponding annihilation operators
in terms of the annihilation operators for linearly polarized photons. 

To fully specify the decomposition in expression in Eq.~(\ref{eq:ef}),
one would need to apply this procedure for each direction $\mathbf{n}$.
However, any decomposition is valid, as physical quantities are independent
of the chosen decomposition, and thus the chosen one is a matter of
convenience.

\bibliographystyle{unsrt}
\bibliography{theory_references}

\end{document}